\documentclass[article]{aa}
\usepackage[varg]{txfonts}
\usepackage{graphicx}
\usepackage{bm}
\usepackage{natbib,twoopt}
\usepackage[breaklinks=true]{hyperref} 
\bibpunct{(}{)}{;}{a}{}{,} 
\makeatletter
\newcommandtwoopt{\citeads}[3][][]{\href{http://adsabs.harvard.edu/abs/#3}%
{\def\hyper@linkstart##1##2{}%
\let\hyper@linkend\@empty\citealp[#1][#2]{#3}}}
\newcommandtwoopt{\citepads}[3][][]{\href{http://adsabs.harvard.edu/abs/#3}%
{\def\hyper@linkstart##1##2{}%
\let\hyper@linkend\@empty\citep[#1][#2]{#3}}}
\newcommandtwoopt{\citetads}[3][][]{\href{http://adsabs.harvard.edu/abs/#3}%
{\def\hyper@linkstart##1##2{}%
\let\hyper@linkend\@empty\citet[#1][#2]{#3}}}
\newcommandtwoopt{\citeyearads}[3][][]%
{\href{http://adsabs.harvard.edu/abs/#3}
{\def\hyper@linkstart##1##2{}%
\let\hyper@linkend\@empty\citeyear[#1][#2]{#3}}}
\makeatother

\newcommand{\be}{\begin{equation}}
\newcommand{\maw}{\mathcal{W}}

\newcommand{\mapp}{\mathcal{P}}
\newcommand{\ee}{\end{equation}}  
\newcommand{\dd}{{\rm d}}
\newcommand{\lp}{\left(}
\newcommand{\rp}{\right)}
\newcommand{\vk}{\textbf{k}}
\newcommand{\vr}{\textbf{r}}

\begin{document}
\title{What can the spatial distribution of galaxy clusters tell about their scaling relations?}
\author{Andr\'es Balaguera-Antol\'{\i}nez}
\institute{Argelander Institute f\"ur Astronomie, Universit\"at Bonn, Auf dem H\"ugel 71, D-53121 Bonn, Germany \\
INAF Osservatorio Astronomico di Roma, Via di Frascati 33, 00040 Monte Porzio Catone, Italy\\
\email{abalan@astro.uni-bonn.de}
}
\date{Recieved /Accepted}
\abstract{The clustering of galaxy clusters is sensitive not only to the parameters characterizing a given cosmological model, but also to the links between cluster intrinsic properties e.g., the X-ray luminosity, X-ray temperature and the total cluster mass. These links, referred to as the \emph{cluster scaling relations}, represent the tip of the iceberg of the so-called cross-roads between cosmology and astrophysics on the cluster scale.}{In this paper we aim to quantify the capability of the inhomogeneous distribution of galaxy clusters, represented by the two-point statistics in Fourier space, to retrieve information on the underlying scaling relations. To that end, we make a case study using the mass X-ray luminosity scaling relation for galaxy clusters and study its impact on the clustering pattern of these objects.} {To characterize the clustering of galaxy clusters, we define the luminosity-weighted power spectrum and introduce the \emph{luminosity power spectrum} as direct assessment of the clustering of the property of interest, in our case, the cluster X-ray luminosity. Using a suite of halo catalogs extracted from $N$-body simulations and realistic estimates of the mass X-ray luminosity relation, we measured these statistics with their corresponding covariance matrices. By carrying out a Fisher matrix analysis, we quantified the content of information (by means of a figure-of merit) encoded in the amplitude, shape, and full shape of our probes for two-point statistics.} {The full shape of the luminosity power spectrum, when analyzed up to scales of $k\sim 0.2\,h\, {\rm Mpc}^{-1}$, yields a figure of merit which is two orders of magnitude above the figure obtained from the unweighted power spectrum, and only one order of magnitude below the value encoded in X-ray luminosity function estimated from the same sample. This is a significant improvement over the analysis developed with the standard (i.e., unweighted) clustering probes. }{The measurements of the clustering of galaxy clusters and its explicit dependence on the cluster intrinsic properties can contribute to improving the degree of knowledge regarding the underlying links between cluster observables and the cluster masses. We therefore suggest future clustering analysis of galaxy clusters to implement the weighted statistics and especially the luminosity (or any other property of interest) power spectrum when aiming at simultaneously constraining cosmological and astrophysical parameters.}

\maketitle

\section{Introduction}
The current understanding of the observed abundance and clustering of galaxy clusters relies on the large-scale statistical properties of dark matter halos \citep[e.g.,][]{1986ApJ...304...15B,jenkins,2003MNRAS.341.1311S,warren,tinker}. Accordingly, the mapping between cluster observables (e.g., X-ray luminosities, X-ray temperatures) and cluster mass is an ingredient of paramount relevance in the process of retrieving cosmological information from galaxy cluster experiments \citep[see, e.g.][]{2003MNRAS.342..163P,2006ApJ...648..956S,2010ApJ...715.1508S,2011ARA&A..49..409A,2012arXiv1210.7303S,2013arXiv1303.5080P}. Such mapping, referred to as \emph{the cluster scaling relations}, represents a simple way of characterizing the complex baryonic processes taking place within galaxy clusters with only few parameters \citep[see, e.g.][]{sarazin, 2010gfe..book.....M,2010ApJ...715.1508S,2012ARA&A..50..353K}. 

The cluster scaling relations are often calibrated by direct measurements of cluster-intrinsic properties and masses \citep[e.g.,][]{2001A&A...368..749F,2006ApJ...648..956S,2009ApJ...703..982G,2009A&A...498..361P,2011A&A...536A..12P}. These usefulness of these measurements strongly depends on factors such as the definition of mass and the dynamical state of galaxy clusters \citep[e.g.,][]{rasia}, that are potentially plagued by systematic effects. In view of this, some studies adopt a  self-calibrated approach in which the set of scaling relations and cosmological parameters are jointly constrained \citep[e.g.,][]{2010MNRAS.406.1773M,2013MNRAS.tmp.1209R} using the cluster abundance as a cosmological probe \citep[e.g.,][]{2002ApJ...566...93B, vik,2011ARA&A..49..409A,2013arXiv1303.5080P}.

Beyond the one-point statistics of galaxy clusters, their spatial distribution, which is often statistically characterized by either the power spectrum $P(k)$ or its Fourier counterpart, itself the two-point correlation function $\xi(r)$ \citep[e.g.,][]{1980lssu.book.....P}, can provide insight into the links between light and matter on cluster scales \citep[e.g.,][]{pillepich}. A simple example of this is the phenomenon of halo-bias, i.e., the increase in the halos clustering strength as a function of the halo mass \citep[e.g.,][]{1986MNRAS.222..323K,1996MNRAS.282..347M,1999MNRAS.308..119S, 2005ApJ...631...41T,pillepich_nm,2013P}. Observationally, this behavior has been detected as a function of intrinsic properties, such as the X-ray luminosity \citep[e.g.,][]{2001A&A...368...86S,2011MNRAS.413..386B}, allowing us to explore the attributes of the scaling relation by means of the \emph{amplitude} of the clustering signal. We refer to this as the \emph{indirect} clustering dependence upon intrinsic properties.  On the other hand, a \emph{direct} scrutiny of the impact of cluster intrinsic properties on the clustering pattern can be achieved by means of the so-called marked (or weighted) statistics \citep[e.g.,][]{2001}. Marked statistics have been applied to galaxy redshift samples \citep[e.g.,][]{2005MNRAS.364..796S,2006MNRAS.369...68S,2009MNRAS.392.1080S,2009MNRAS.395.2381W} as an attempt to explore the environmental dependence and clustering of a given intrinsic property of galaxies, such as luminosities, stellar masses and colors.

In this paper we quantify the ability of the three-dimensional clustering signal from galaxy clusters to retrieve information on the underlying cluster scaling relations, represented here by the link between the cluster masses and X-ray luminosities. To this end, we measure the two-point marked-statistics in Fourier space by means of the luminosity-weighted power spectrum and the luminosity power spectrum. The latter is introduced \---to our knowledge\--- for the first time in this paper, as an attempt to explicitly study the X-ray luminosity dependence of the clustering of galaxy clusters. 
We use a suite of halo catalogs built from $N$-body simulations, together with a realistic estimate of cluster scaling relations, and carried out a Fisher-matrix analysis to determine the sensitivity of these clustering probes 
to the parameters characterizing the scaling relation. We stress that our analysis is not a forecast for a particular experiment in view of a given cosmological model. Instead, we present a comparison between different clustering-related probes that can be implemented in forthcoming galaxy cluster samples in order to extract astrophysical and cosmological information. Our findings suggest that the direct assessment of the clustering of X-ray luminosity (or other physical properties) can help establish tight constraints on the cluster scaling relations. 

The outline of this paper is as follows. In Section $2$ we introduce the suite of $N$-body simulations and the scaling relations used to assign observables to the dark matter halos. We introduce the cluster luminosity-weighted power spectrum and the luminosity power spectrum and define a set of observables from which the information on the scaling relation is retrieved. In Section~\ref{sec:fish} we evaluate the information content in those observables. We summarize our conclusions in Section \ref{sec:conc}.

\section{Probes for cluster scaling relations}\label{probes}
\subsection{Halo catalogs}\label{sec:sim}
We base our analysis on the L-BASICC $N$-body simulations \citep[][]{2008MNRAS.383..755A} represented by a suite of $N=50$ realizations of the same flat $\Lambda$CDM cosmological model at redshift zero. The simulations are characterized by a matter density parameter $\Omega_{\rm mat}=0.237$, a baryon density parameter $\Omega_{\rm ba}=0.046$, a dimensionless Hubble parameter $h=0.73$\footnote{The Hubble constant $H_{0}$ in units of
$100 \,{\rm km}\,{\rm s}^{-1}{\rm Mpc}^{-1}$.}, linear rms mass fluctuations within $8\,h^{-1}\,{\rm Mpc}$ of $\sigma_{8}=0.773$, and a scalar spectral index $n_{s}=0.997$, following the evolution of $448^{3}$ dark matter particles in a comoving box with volume $V=(1340\,\,h^{-1}\,{\rm Mpc})^{3}$. Halo catalogs were built by means of a friends-of-friends algorithm, characterized by a linking length of $0.2$ times the mean interparticle separation and a minimum mass of $1.73\times 10^{13}\,h^{-1}M_{\odot}$ (which corresponds to ten dark matter particles).

\subsection{Cluster scaling relation}\label{sec:sim_l}
We assign X-ray luminosities\footnote{If not explicitly written, X-ray luminosities and masses are expressed in units of $10^{44}h^{-2} {\rm erg}\,s^{-1}\,h^{-2}$ and $10^{14}\, h^{-1}M_{\odot}$ respectively.} $L$ to dark matter halos by means of a log-normal conditional probability distribution $\mapp(L|M)\dd L$ (hereafter scaling relation), specifying the probability that a cluster has an X-ray luminosity in the range $L, L+\dd L$, conditional on its mass $M$:
\be\label{sr}
\mapp(L|M)\dd L=\frac{1}{\sqrt{2\pi}\tilde{\sigma}}{\rm exp}\left[-\frac{1}{2\tilde{\sigma}^{2}}\lp\ell-\langle \ell|M\rangle \rp^{2}\right]\dd \ell,
\ee
 where $\ell\equiv \log \lp  L/(10^{44}{\rm erg}\,{\rm s}^{-1}\,h^{-2})\rp$ and $\tilde{\sigma}$ denotes the intrinsic scatter in $\ell$ at a fixed mass scale $M$, i.e., $\tilde{\sigma}^{2}=\langle \ell^{2}|M \rangle- \langle \ell|M \rangle^{2}$. For the mean of the scaling relation we use a power law with amplitude $\alpha$ and slope $\gamma$,
\be
\langle \ell|M\rangle =\alpha +\gamma \log\lp \frac{M}{10^{14}h^{-1}M_{\odot}}\rp.
\ee
As fiducial values we use $\alpha=-0.64$, $\gamma=1.27$ and $\sigma=0.15$, with $\sigma=\tilde{\sigma}/\ln(10)$. These numbers follow from realistic estimates of the cluster's mass-X ray luminosity relation \citep[e.g.,][]{2012MNRAS.425.2244B}. However, the exact figures are not relevant for the purposes of this work, as long as we are not constructing galaxy-cluster catalogs constrained to follow a particular selection function. The Fisher matrix analysis presented in Sect.~\ref{sec:fish} is be dedicated to the sensitivity of a set of observables\--- to be defined in Sect.~\ref{sec:meas}\--- to the set of parameters $\{\mathcal{X}\}\equiv \{\alpha,\gamma,\sigma\}$.

For the forthcoming analysis, we consider objects with luminosities within the range $3\times 10^{42}\leq L/(h^{-2}\,{\rm erg}\,s^{-1})\leq 5\times 10^{44}$, which leads to realizations with $N_{\rm cl}\sim 4\times 10^{5}$ entries. To study the dependence of clustering with luminosity, we split this range in $n_{\ell}=10$ disjoint and equally log-spaced bins. When the luminosity dependence is not explicitly shown in the forthcoming expressions, it is assumed that clusters with luminosities in the full luminosity range have been considered for the analysis.

\subsection{Clustering estimates}\label{sec:meas}
\subsubsection{Definitions}
Let $n_{\rm w}(\vr;L)$ denote the number density of weighted (or marked) halos at a position $\vr$ and luminosity $L$ and $\bar{n}_{w}(L) $ its mean value. We aim to measure the luminosity-weighted power spectrum by defining the fluctuation $\delta_{\rm w}(\vr;L)\equiv (n_{w}(\vr;L)-\bar{n}_{w}(L))/\bar{n}_{w}(L)$. Similarly, let $\delta_{\rm h}(\vr;L)\equiv (n(\vr;L)-\bar{n}(L))/\bar{n}(L)$ denotes the unmarked cluster fluctuation, with $\bar{n}(L)$ as the mean number density of clusters with luminosity $L$. As weight we use the cluster X-ray luminosity, $w_{i}\equiv L_{i}/\bar{L}$, where $\bar{L}$ is the first sample moment (or mean) of the luminosity in the sample. By definition $\bar{w}=1$, which in turn implies that $\bar{n}_{w}=\bar{w}\,\bar{n}=\bar{n}$. The quantity $n_{w}(\vr;L)$ denotes an inhomogeneous marked point process, in which two sources of stochasticity are present, namely, the associated to the sampling process giving rise to the observed $n(\vr;L)$ and the one due to the stochastic nature of the marks. We assume that the halo distribution $n(\vr;L)$ is the result of an inhomogeneous Poisson point process. Strictly speaking, a Poisson model cannot properly describe the distribution of galaxy clusters since these objects are subject to the so-called halo exclusion effects \citep[e.g.,][]{2002ApJ...565...24P, 2005ApJ...631...41T,2007PhRvD..75f3512S}. This effect arises because halos are treated as disjointed entities that are not allowed to overlap, leading to a correlation function $\xi(r)\to -1$ on scales below the minimum scale probed by the halo catalog (for an ideal sample of spherical halos, this scale equals twice the radius of the smaller halo \citep[e.g.,][]{1980lssu.book.....P}). We discuss this subject briefly in Appendix~\ref{exclusion}.

\begin{figure}
\center
\includegraphics[width=9cm, angle=0]{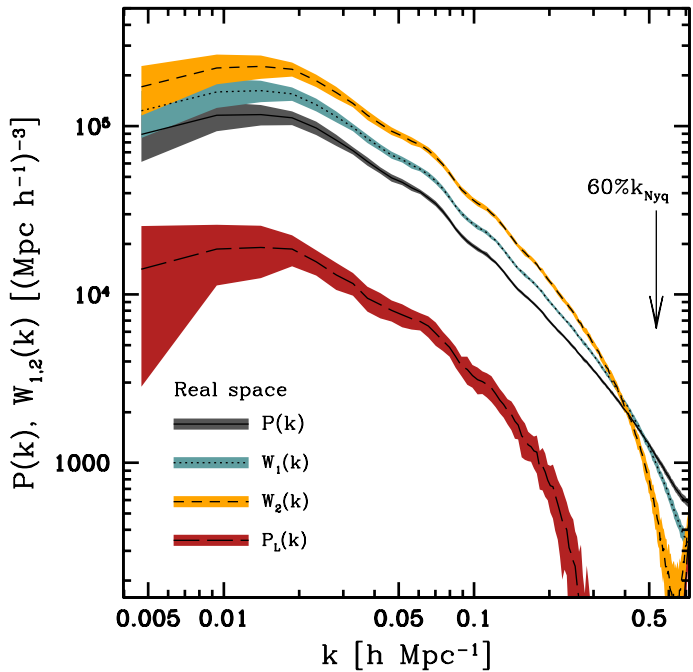}
\caption{Mean power spectra $\hat{P}(k)$, $\hat{\maw}_{1,2}(k)$ and $\hat{P}_{L}(k)$ (defined in Sect.~\ref{sec:meas}) obtained from the ensemble of $N$-body simulations for galaxy clusters within the full luminosity range as described in the text. The $60\%$ of the Nyquist frequency is marked with an arrow. The shaded regions denote the standard deviation obtained from the ensemble of $N$-body simulations.}\label{redis0}
\end{figure}

\subsubsection{Probes of clustering}
The weighted halos are embedded into a $380^{3}$ cubic grid using a cloud-in-cell mass assignment scheme \citep{1988csup.book.....H}. We use the FFTW algorithm \citep{2012ascl.soft01015F} to compute the Fourier transform of $\delta_{\rm w}(\vr;L)$, $\tilde{\delta}_{w}(\vk;L)$, correcting thereafter for aliasing effects \citep[see e.g.,][]{2008MNRAS.383..755A}. We next obtain estimates\footnote{We denote by $\hat{f}$ the estimates of the quantity $f$.} of the unmarked $P(k;L)$, the cross power spectrum between the unweighted and weighted halo density fields $\maw_{1}(k;L)$ and the luminosity-weighted power spectrum $\maw_{2}(k;L)$ by 
\be\label{pp}
\hat{P}(k_{i};L_{j})=\langle |\tilde{\delta}_{\rm h}(\vk;L_{j})|^{2}\rangle_{k_{i}}-\frac{1}{\bar{n}(L_{j})},
\ee
\be\label{w1}
\hat{\maw}_{1}(k_{i};L_{j})=\langle {\mathcal Re}(\tilde{\delta}_{\rm h}(\vk;L_{j})\tilde{\delta}_{\rm w}^{*}(\vk;L_{j})) \rangle_{k_{i}}-\frac{\overline{w}_{j}}{\bar{n}(L_{j})},
\ee
\be\label{w2}
\hat{\maw}_{2}(k_{i};L_{j})=\langle |\tilde{\delta}_{\rm w}(\vk;L_{j})|^{2}\rangle_{k_{i}}-\frac{\overline{w^{2}}_{j}}{\bar{n}(L_{j})},
\ee
where $\langle \cdot \rangle_{k_{i}}$ denotes average in spherical shells of width $\Delta k=2\pi V^{-1/3}$ centered at $k_{i}$. The second term in Eq.~(\ref{pp}) corresponds to the Poisson shot-noise correction \citep[e.g.,][]{1980lssu.book.....P}, an extra variance of the halo field induced by discreetness. Similarly, the extra variance in Eqs.~(\ref{w1}) and (\ref{w2}) are the product of the variance related to the Poisson point-like process and the one induced by the luminosity distribution. The term $\overline{w^{2}}_{j}$ denotes the second sample moment of the weights within the $j^{\rm th}$ luminosity bin. Again that in Eq.~(\ref{w1}), $\overline{w}_{j}=1$.

\begin{figure}
\center
\includegraphics[width=9cm, angle=0]{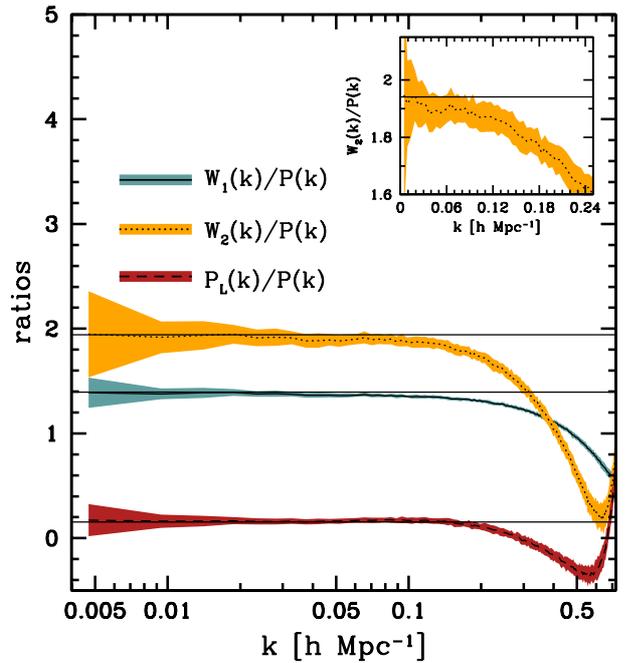}
\caption{Ratio between the spectra $\hat{\maw}_{1,2}(k)$ and $\hat{P}_{L}(k)$ to the cluster power spectrum $\hat{P}(k)$, shown in real space. The horizontal lines show the predictions presented in Appendix~\ref{apl}. The shaded regions denote the standard deviation obtained from the ensemble of $N$-body simulations. The inset shows a close-up of the ratio $\maw_{2}/P(k)$, showing a clear scale-dependent trend from scales  $k\gtrsim 0.06 \,h\,{\rm Mpc}^{-1}$. }\label{redis0r}
\end{figure}
Along with the measurements in real space, we also measured two-point statistics in redshift space by means of the distant-observer approximation, i.e., shifting the position of dark matter halos along the $x-$axis $x\to x+v_{x}/H_{0}$ where $v_{x}$ is the $x-$ component of the peculiar velocity of the halo center of mass. We keep Fourier modes up to a $\sim 60\%$ of the Nyquist frequency $k_{\rm Ny}=0.89\,h\,{\rm Mpc}^{-1}$, where the corrections due to the assignment scheme are accurate \citep[e.g.,][]{2005MNRAS.362..505C}.

In Fig.~\ref{redis0} we show the mean of the real-space power spectra $\hat{P}(k)$ and $\hat{\maw}_{1,2}(k)$ (computed using the full luminosity range) with the corresponding standard deviation obtained from the ensemble of realizations of the L-BASICC simulation. At first glance, the estimates of marked power spectra $\hat{\maw}_{1,2}(k)$ behave as scaled versions of $\hat{P}(k)$. Indeed, on large scales ($k\lesssim 0.04\,h\,{\rm Mpc}^{-1}$), the ratios $\maw_{1,2}(k)/P(k)$ are described well by a constant factor, as explicitly shown in Fig.~\ref{redis0r}. This suggests that the amplitude of the marked power spectra $\hat{\maw}_{1,2}(k)$ can be explained, as a first approximation, by including the information of the scaling relation within a sort of luminosity bias. Indeed, the theoretical expectations described in Appendix~\ref{apl} and represented in Fig.~\ref{redis0r} by the solid lines show that this is the case. 

A more careful analysis reveals, though, a scale dependence of the ratios $\maw_{1,2}(k)/P(k)$, which is statistically significant  even on large and intermediate scales $k\gtrsim 0.04 \,h\,{\rm Mpc}^{-1}$ (also present in redshift space). This is shown in the inset of Fig.~\ref{redis0r}. We note that $\hat{\maw}_{1,2}(k)$ and $\hat{P}(k)$ are equally affected by the nonlinear evolution of the underlying matter density field and the exclusion effect (i.e., they are estimated from the same sample and thus the smallest populated bin of separation is the same, albeit differently weighted), as well as by the same scale-dependent halo-mass bias \citep[see, e.g.,][]{2005ApJ...631...41T}. Therefore, the scale dependency observed in the ratios $\maw_{1,2}(k)/P(k)$ can be regarded as a signature of the distribution of the weights $w_{i}w_{j}$ within pairs at different separations, suggesting that the shape of the luminosity-weighted power spectrum is sensitive to the scaling relation $\mapp(L|M)$. This is explored in Sect.~\ref{sec:fish}.

\subsubsection{Introducing a new probe}
As an attempt to isolate the signal of the clustering of the X-ray luminosity from the unmarked power spectrum, we define $\delta_{L}(\vr;L)\equiv \delta_{\rm w}(\vr;L)-\delta_{\rm h}(\vr;L)=(n_{w}(\vr;L)-n(\vr;L))/\bar{n}(L)$.
 This fluctuation has zero mean, $\langle \tilde{\delta}_{L}(\vk;L)\rangle=0$, and variance $P_{L}(k;L)\equiv \langle |\tilde{\delta}_{L}(\vk;L)|^{2}\rangle$, where $\langle \cdot \rangle$ denotes an average over an ideal ensemble of realizations. We refer to this variance as the \emph{luminosity power spectrum}. Thus defined, the variable $\delta_{L}$ accounts for the fluctuations in the weighted number density with respect to the unweighted number density. Since by construction both number densities share the same mean value, this simple subtraction avoids us from having to assume that the observed mean number density is the true mean density, and thus, the estimates of the luminosity power spectrum are free of the so-called integral constraint \citep[e.g.,][]{1991MNRAS.253..307P}. Since the mean number density $\bar{n}$ only enters into the definition of $\delta_{L}$ as a normalization, the possible difference between the observed and true number density would equally affect all Fourier modes in $P_{L}(k;L)$ by some constant factor. 
The estimates of the luminosity power spectrum are obtained following the same procedure as described in the previous section, namely, a shell-average of the estimate $|\tilde{\delta}(\vk;L)^{2}|$ with a shot-noise correction $S_{L}$:
\be\label{lal}
\hat{P}_{L}(k_{i};L_{j})=\langle |\tilde{\delta}_{L}(\vk;L_{j})|^{2}\rangle_{k_{i}}-S_{L}.
\ee
According to the definition of $\delta_{L}(\vr)$, the luminosity power spectrum can be written as the combination of the luminosity-weighted power spectra $P_{L}=\maw_{2}-2\maw_{1}+P$. Thus, we can identify the shot-noise correction with $S_{L}=\sigma^{2}_{w\,j}/\bar{n}(L_{j})$, where $\sigma^{2}_{w\,j}=\overline{w^{2}}_{j}-1$. Figure~\ref{redis0} shows the mean of the estimates of $P_{L}(k)$ as determined from the ensemble of $N$-body simulations (in the full luminosity range), together with its standard deviation. As shown in Fig.~\ref{redis0r}, on large scales the shape of $\hat{P}_{L}(k)$ can be understood as a biased halo power spectrum, i.e., $P_{L}(k)/P(k)\sim {\rm constant}$, which is even prevailing on intermediate scales ($\sim 0.1  \,h\,{\rm Mpc}^{-1}$). In view of the results shown in Appendix~\ref{apl}, the luminosity power spectrum can be thought of as directly probing the mean scaling relation. 

Table~\ref{table} summarizes the definitions of the different power spectra introduced in this section.

 \subsection{Two-point statistics as probe for cluster scaling relations}
 Generally, the information related to the parameters of a model explaining the observed two-point statistics can be extracted in three different ways: $i)$  using the amplitude of the clustering signal, $ii)$ using its full shape \cite[e.g.,][]{2012MNRAS.425..415S}, or $iii)$ isolating particular features such as the position of the baryonic acoustic peak in the correlation function \citep[e.g.,][]{2012MNRAS.427.3435A, 2012ApJ...749...81H} or the turn-over in the power spectrum \citep[e.g.,][]{2010MNRAS.404...60R,2013MNRAS.429.1902P}. We study the information content in cases $i)$ and $ii)$, and leave the assessment of the dependence of particular features for future studies, e.g., the acoustic peak in the correlation function, on the cluster scaling relations.

\subsubsection{Amplitude}\label{sec:amp}
The information contained in the amplitude of the clustering signal can be extracted by defining ratios of spectra by the underlying matter power spectrum or in a less model-dependent fashion, by the clustering of the same population of objects at a given luminosity \citep[e.g.,][]{2002MNRAS.332..827N}. As suggested by panel (b) of Fig.~\ref{redis0}, such ratios are approximately constant on large scales (e.g., $k\lesssim 0.04\,h\,{\rm Mpc}^{-1}$). We define for each realization $q=1,\cdots, N$, the following ratios:
\begin{eqnarray}\label{r1}
r^{(q)}_{1}(k_{i};L_{j})\equiv  \lp\frac{\hat{P}^{(q)}(k_{i};L_{j})}{\hat{P}^{(q)}(k_{i};L_{\rm ref})}\rp^{\frac{1}{2}},&& \hspace{-0.5cm}
r^{(q)}_{2}(k_{i};L_{j})\equiv  \lp\frac{\hat{\maw}^{(q)}_{1}(k_{i};L_{j})}{\hat{\maw}^{(q)}_{1}(k_{i};L_{\rm ref})}\rp^{\frac{1}{2}}, \nonumber \\
r^{(q)}_{3}(k_{i};L_{j})\equiv  \lp\frac{\hat{\maw}^{(q)}_{2}(k_{i};L_{j})}{\hat{\maw}^{(q)}_{2}(k_{i};L_{\rm ref})}\rp^{\frac{1}{2}},&& \hspace{-0.5cm}
r^{(q)}_{4}(k_{i};L_{j})\equiv  \lp \frac{\hat{\maw}^{(q)}_{1}(k_{i};L_{j})}{\hat{P}^{(q)}(k_{i};L_{j})}\rp^{\frac{1}{2}}, \nonumber \\
r^{(q)}_{5}(k_{i};L_{j})\equiv   \lp \frac{\hat{\maw}^{(q)}_{2}(k_{i};L_{j})}{\hat{P}^{(q)}(k_{i};L_{j})}\rp^{\frac{1}{2}},&&\hspace{-0.5cm}
r^{(q)}_{6}(k_{i};L_{j})\equiv   \lp \frac{\hat{P}^{(q)}_{L}(k_{i};L_{j})}{\hat{P}^{(q)}_{L}(k_{i};L_{\rm ref})}\rp^{\frac{1}{2}}, \nonumber \\
r^{(q)}_{7}(k_{i};L_{j})\equiv   \lp \frac{\hat{P}^{(q)}_{L}(k_{i};L_{j})}{\hat{P}^{(q)}(k_{i};L_{j})}\rp^{\frac{1}{2}}.&&\hspace{-0.5cm} 
\end{eqnarray}
Our set of large-scale luminosity-dependent observables is then defined as $r^{q}_{\nu}(L_{j})\equiv \langle r^{q}_{\nu}(k_{i};L_{j})\rangle_{\delta k}$, where $\langle\cdot\rangle_{\delta k}$ denotes the average over wavenumbers in the range $\delta k=[0.01, 0.04]\,h\,{\rm Mpc}^{-1}$. The first three probes $r_{1,2,3}(L)$ are defined as ratios of a given power spectrum in different luminosity bins with respect to the same statistics, the latter measured in a fixed (reference) luminosity bin $L_{\rm ref}$, thereby providing estimates of \emph{relative biases} of the marked spectra. The ratios $r_{4,5}(L)$ are instead defined by the unmarked power spectrum in the same luminosity bin, and we refer to these as estimates of \emph{absolute bias}\footnote{We mean absolute with respect to the unmarked cluster power spectrum, not to the matter power spectrum.}. Similarly, the ratios $r_{6,7}(L)$ are estimates of relative and absolute bias extracted from the luminosity power spectrum. In these expressions $L_{\rm ref}$ is taken as containing the value $L_{\star}=0.63\times 10^{44}{\rm erg}\,{\rm s}^{-1}h^{-2}$, the typical X-ray luminosity of galaxy clusters up to redshift $z\sim 0.2$ \citep{2012MNRAS.425.2244B}. We checked that our results do not change substantially when other values of $L_{\rm ref}$ are used, except for the high-luminosity bins, wherein the clustering estimates display a low signal-to-noise ratio. In Appendix~\ref{apl} we show the theoretical predictions for these ratios.

\begin{table*}[t]
\center
\begin{tabular}{|l|l|l|} \hline
Symbol & Definition & Description  \\ \hline
&&\\ 
$w$ & $L /\bar{L}$ & Cluster luminosity in units of the mean luminosity of the sample \\
$n(\vr;L)$ ($n_{w}(\vr)$)  &  & Unweighted (weighted) number density of clusters with luminosity $L$\\
$\delta(\vr;L)$  & $(n(\vr;L)-\bar{n}(L))/\bar{n}(L)$ &Standard (i.e. unweighted) galaxy cluster fluctuation\\
$\delta_{w}(\vr;L)$  & $(n_{w}(\vr;L)-\bar{n})/\bar{n}(L)$ &Weighted galaxy cluster fluctuation\\
$\delta_{L}(\vr;L)$  & $(n_{w}(\vr;L)-n(\vr;L))/\bar{n}(L)$ & Luminosity galaxy cluster fluctuation\\
$\tilde{\delta}(\vk;L)$  &  &Galaxy cluster fluctuation in Fourier space\\
 $\hat{P}(k;L_{j})$ & $\langle |\tilde{\delta}(\vk;L_{j})|^{2}\rangle_{k_{i}}-\frac{1}{\bar{n}(L_{j})}$& 
 Estimates of (shot-noise $S$ subtracted) power spectrum of \\
& & objects in the $j-$the luminosity bin \\
& & averaged in Fourier space within a spherical shell centered at $k_{i}$ \\

 $\hat{\maw}_{1}(k_{i};L_{j})$ & $\langle {\mathcal Re}(\tilde{\delta}_{w}(\vk;L_{j})\tilde{\delta}(\vk;L_{j})) \rangle_{k_{i}}-\frac{\bar{w}}{\bar{n}(L_{j})}$ &Cross (i.e, weighted-unweighted) power spectrum \\
 $\hat{\maw}_{2}(k_{i};L_{j})$ &$\langle |\tilde{\delta}_{w}(\vk;L_{j})|^{2}\rangle_{k_{i}}-\frac{\bar{w^{2}}}{\bar{n}(L_{j})}$ &Weighted power spectrum\\
 $\hat{P}_{L}(k_{i};L_{j})$ & $\langle |\tilde{\delta}_{L}(\vk;L)|^{2}\rangle_{k_{i}}-\frac{\sigma^{2}}{\bar{n}(L_{j})}$ &Luminosity power spectrum  \\
\hline

\hline 
\end{tabular}\caption{\label{tableres} Different power spectra defined in the text. The symbol $\langle\cdot \rangle_{k_{i}}$ denotes average in Fourier space.}
\label{table}
\end{table*}

\subsubsection{Full shape}\label{sec:pl}
The measurements of the luminosity-weighted power spectrum shown in Sect.~\ref{sec:meas} revealed that the scaling relation $\mapp(L|M)$ not only affects the amplitude of the weighted power spectra $\maw_{1,2}(k)$, but can also play a role in shaping their broad-band signal. To quantify the information encoded in the full shape of the measured spectra $P(k)$, $\maw_{1,2}(k)$, and $P_{L}(k)$, we use their estimates in the range of wave numbers $0.02\leq k/(h\,{\rm Mpc}^{-1})\leq 0.2$, with $\tilde{N}_{k}=32$ Fourier modes, and use the full X-ray luminosity range. 

As pointed out in Sect.~\ref{sec:meas}, the variance of the fluctuations of the luminosity-weighted and luminosity power spectrum has a contribution that embodies two forms of randomness: one associated to the finite number of halos, the other to the distribution of the marks compared to its mean. These two effects are represented by Eq.(\ref{lal}). The shot-noise subtraction in the power spectrum of galaxy clusters is a delicate issue. Clusters, as rare objects, display number densities that are low enough to generate a shot-noise contribution that can be traceable even on the largest scales. As an example of its relevance, it can be seen that estimates of power spectra without shot-noise correction can generate a scale-dependent bias on scales where the ratios between corrected spectra were fairly described by a constant factor \citep[see, e.g.,][]{2012MNRAS.420.3469P}. Strictly speaking, not subtracting the shot-noise correction from the estimates of power spectra is not an off base procedure, given that this correction is a model in itself and as such can be included in the modeling of the clustering signal. To asses the impact of the shot-noise correction in the amount of information we can retrieve, we consider below the case when the information content from the luminosity power spectrum is extracted from its shot-noise uncorrected estimate. 

\section{Results: information content}\label{sec:fish}
In this section we carry out a Fisher matrix analysis \citep[see, e.g.,][]{1997ApJ...480...22T} in order to quantify the amount of information on the mass-X ray luminosity scaling relation that our clustering-related observables contain. Before proceeding, we recall that in the context of galaxy cluster experiments, one-point statistics \---such as cluster number counts\--- represent the standard probe for cosmological parameters and scaling relations \citep[e.g.,][]{2004ApJ...613...41M, lima,2010PhRvD..81h3509C,pillepich}. To compare our results \---which are based on two-point statistics\--- with a one-point statistics probe, we use the X-ray luminosity function $\hat{\Phi}(L)$ (XLF hereafter) measured in $\tilde{n}_{\ell}=20$ equally log-spaced bins. Given that the cosmological parameters are kept fixed throughout the analysis\footnote{Even though the dependence on cosmological parameters of the cluster number counts and luminosity function is slightly different, their dependence on the parameters of the scaling relation is the same.}, the information content in the XLF can provide a fair level of comparison to assess the capability of the clustering probes in retrieving information on the cluster scaling relation against the standard procedures. Our set of observables is summarized as
\be\label{muq}
 \mu^{q}_{j}=\{r_{\nu=1,\cdots,7}(L_{j}),\hat{P}(k_{j}),\hat{\maw}_{\nu=1,2}(k_{j}),\hat{P}_{L}(k_{j}),\hat{\Phi}(L_{j})\},
\ee
where the index $j$ runs from $(1,\cdots, n_{\ell})$ for the ratios $r_{\nu}(L)$,  $(1,\cdots,\tilde{N}_{k})$ for the power spectra, and $(1,\cdots,\tilde{n}_{\ell})$ for the XLF. For each observable, its mean $\bar{\bm{\mu}}$ and covariance matrix $\bm{C}$ are given by $\bar{\mu}_{i}=\langle \mu_{i} \rangle_{\rm ens}$ and $C_{ij}=\langle \mu_{i}\mu_{j}\rangle_{\rm ens}-\bar{\mu}_{i}\bar{\mu}_{j}$ respectively, where $\langle \cdot\rangle_{\rm ens}$ denotes averages over the ensemble of realizations of the L-BASICC simulations.

\begin{figure}
\center
\includegraphics[width=8.5cm, angle=0]{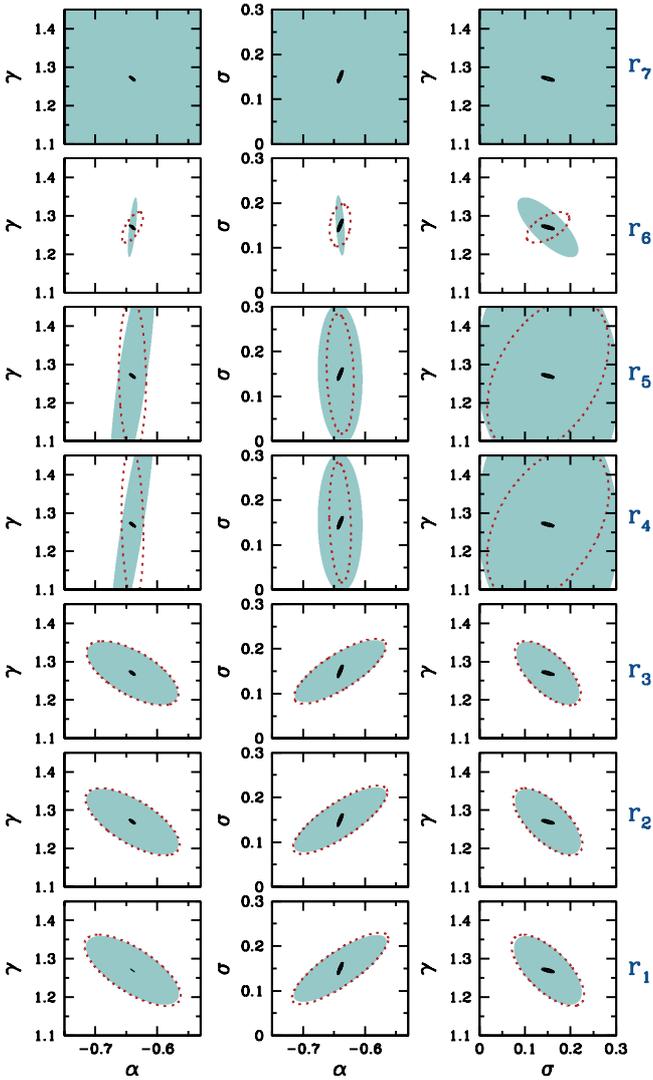}
\caption{Joint $1\sigma$ error ellipses for the parameters $\{\alpha,\gamma,\sigma\}$ of the scaling relation $\mapp(L|M)$, obtained from the Fisher-matrix analysis in real (shaded area) and redshift space (dotted lines contours), using the set $r_{\nu=1,\cdots, 7}(L)$ defined in Eq.~(\ref{r1}) as observables. The small filled contours represent the results using the XLF. }\label{red2}
\end{figure}

The Fisher matrix represents the ensemble average of the Hessian of the natural logarithm of the likelihood function of a data set given a model, and its customary application aims to forecast uncertainties of a set of model parameters from a given experiment. Our approach is slightly different, though, since we aim to quantify the sensitivity of each of the observables defined in Eq.~(\ref{muq}) to the set of parameters $\{\mathcal{X}\}$ (defined in Sect~\ref{sec:sim_l}) using measurements and covariance matrices directly extracted from the simulations. In other words, we assume perfect knowledge of model and the covariance for our observables.
Under the assumption that the observables are drawn from a Gaussian distribution, the elements of the Fisher matrix $\bm{F}$ are written as
\be\label{fis}
F_{x y}=\frac{1}{2}{\rm Tr}\left[\bm{C}^{-1}\bm{C}_{,x}\bm{C}^{-1}\bm{C}_{,y}+\bm{C}^{-1}\lp \bar{\bm{\mu}}_{,x}\bar{\bm{\mu}}_{,y}^{T}+\bar{\bm{\mu}}_{,y}\bar{\bm{\mu}}_{,x}^{T} \rp \right]|_{x=x_{\rm fid}}^{y=y_{\rm fid}},
\ee
where $,x \equiv \partial/\partial x$ and $x_{\rm fid}$ denotes the fiducial values of the parameters $\{\mathcal{X}\}$, defined in Sect.~\ref{sec:sim_l}. The covariance matrix of the parameters of the model is $C_{xy}\approx F_{xy}^{-1}$, and the marginal error of each parameter is obtained from the Cram\'er-Rao inequality $\sigma_{x}\geq (\bm{F}^{-1})_{xx}^{1/2}$. We used a double-side variation to accurately estimate the derivatives with respect to the parameters. This leads us to computing $27$(models) $\times 50$(realizations) $\times 2$(real and redshift-space estimates)$\times 3$(marked spectra)$\times 10$ (luminosity bins) $=8.1\times 10^{5}$ estimates of power spectrum.

Three key aspects of our Fisher matrix analysis are taken into account. Firstly, by means of the so-called D’Agostino K2 goodness-of-fit test, we have verified that the distribution of the observables defined in Eq.~(\ref{muq}) within the suite of realizations of the $N$-body simulations is compatible to $95$ per cent confidence with a normal distribution within the ranges of the X-ray luminosity and wavenumbers of interest. Secondly, we note that when computing the Fisher matrix for all our observables (with mean values $\bar{\bm{\mu}}\neq 0$), we omit the term containing the derivatives of the covariance matrix, since it adds spurious information that shrinks the marginalized errors considerably \citep{2013A&A...551A..88C}. Finally, when exploring the information content encoded solely in the shape of the power spectra, we treat the amplitude of the mean spectra as a nuisance parameter. Under the assumption that these amplitudes have flat or Gaussian-distributed priors, analytical marginalization of the covariance matrix over these parameters is possible \citep[e.g.,][]{2002PhRvD..66j3511L,2010MNRAS.408..865T}. For a flat prior on these amplitudes, the inverse of the marginalized covariance matrix is obtained by subtracting the factor $\lp \bm{C}^{-1}\bar{\bm{\mu}}^{T}\bar{\bm{\mu}}\rp/\lp{\rm Tr}[\bar{\bm{\mu}}{\bm C}\bar{\bm{\mu}}^{T}]\rp$ from the original inverted covariance matrix.

\begin{figure*}
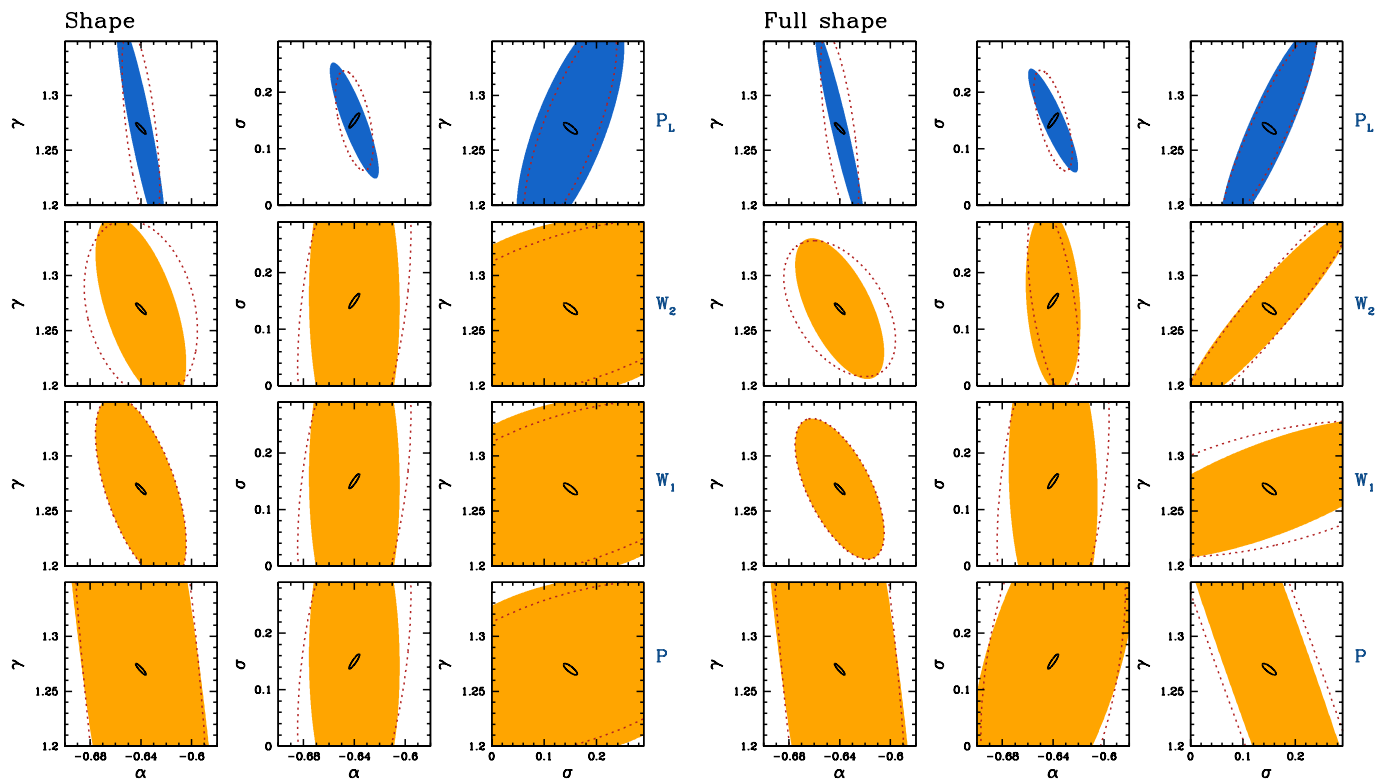

\center
\includegraphics[width=9.1cm, angle=0]{ellipses_power_margi1a.epsi}
\includegraphics[width=9.1cm, angle=0]{ellipses_power1a.epsi}
\caption{Joint $1\sigma$ error ellipses for the parameters $\{\alpha,\gamma,\sigma\}$ of the scaling relation $\mapp(L|M)$, obtained from the Fisher-matrix analysis of power spectra $P(k)$, $\maw_{1,2}(k)$ and $P_{L}(k)$ defined in Sect.~\ref{sec:meas}. Results are shown in real (shaded area) and redshift space (dotted lines). The estimates of these spectra are obtained within the full-luminosity range and using Fourier modes in the interval $0.02<k/(h\,{\rm Mpc}^{-1})<0.2$. The solid contours represent the results using the XLF. Right panel shows the results using after marginalizing over the amplitude of the power spectra. Left panel shows the results using the full-shape. The color of the first row is meant to highlight the results obtained with the luminosity power spectrum.}\label{red22}
\end{figure*}

We quantify the information obtained from each of our observables by a figure-of-merit (FoM hereafter) defined as the inverse of the area of the marginalized $68\%$ error ellipse of each pair of parameters. We also compute the total FoM given by the volume of the three-dimensional $68\%$ error ellipsoid.  We present the results from the Fisher matrix analysis in Figs.~\ref{red2} and \ref{red22}. Figure \ref{rex} condenses the FoM derived from our set of observables. We draw our conclusions based on these three figures.

\subsection{Information content in the amplitude}

Figure~\ref{red2} shows the joint $1\sigma$ error ellipses obtained from the Fisher-matrix analysis of the set $r_{\nu=1,\cdots,7}(L)$. The results are presented in real and redshift space. We next draw some conclusions based on this figure, in conjunction with Fig.~\ref{rex}.
\begin{itemize}
\item The sensitivity to the parameters of the scaling relation varies little between the observables $r_{1,2,3}(L)$. This is also true when these observables are measured in redshift space. The estimates associated to absolute biases $r_{4,5}(L)$ are slightly more sensitive to the signal of the spectra in redshift space.
\item When the ratios are defined by the unmarked power spectrum (i.e., $r_{4,5}(L)$), the marginalized errors on the amplitude, $\sigma_{\alpha}$,  decrease while $\sigma_{\gamma}$ increases, with a net decrease in the total FoM. Therefore, the set $r_{1,2,3}(L)$ proves to be more sensitive to the set $\{\gamma,\sigma\}$, while the set $r_{4,5}(L)$ sets tighter constraints on amplitude $\alpha$.  
\item The estimates of relative bias extracted from the luminosity power spectrum $r_{6}(L)$ generate FoM that are approximately two orders of magnitude above those obtained from the luminosity weighted power spectrum, say $r_{3}(L)$, marking a noticeable improvement. The total FoM of derived from this probe is only $\sim 40$ times below the one derived from the XLF. On the other hand, the estimate of absolute bias $r_{7}(L)$ sets poor constraints. 
\item In general, better constraints are obtained from the estimates of relative bias. As pointed out before, the selection of the reference luminosity does not affect our conclusions, except for very high values of $L_{\rm ref}$. 
\end{itemize}
We conclude that, regarding the amplitude of the cluster power spectrum, the unmarked analysis is accurate enough and no extra-information is gained by measuring the luminosity dependence of the amplitude of the luminosity-weighted power spectrum. On the other hand, we have shown that the estimates of luminosity bias obtained from the luminosity power spectrum, $r_{6}(L)$, allows us to retrieve an amount of information on the scaling relation that is only one order of magnitude below the one characterizing the XLF.

\subsection{Information content in the full shape}
\begin{figure}
\center
\includegraphics[width=8cm, angle=0]{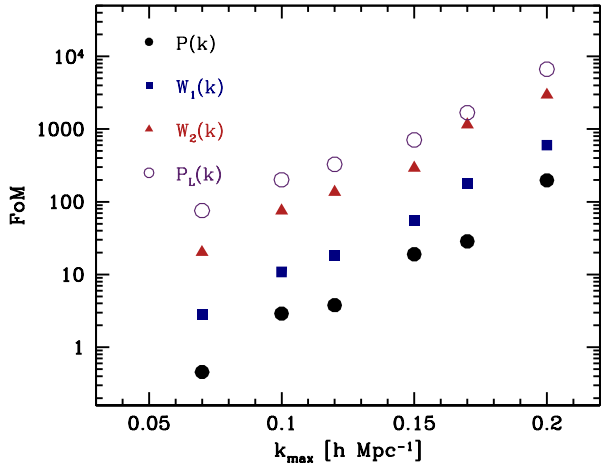}
\caption{FoM obtained from the $1\sigma$ error ellipsoids of the parameters of the scaling relation as a function of the maximum wavenumber $k_{\rm max}$. The FoM shown correspond to the information contained in the full shape of the clustering probes.}
\label{kmax}
\end{figure}

\begin{figure}
\center
\includegraphics[width=8.7cm, angle=0]{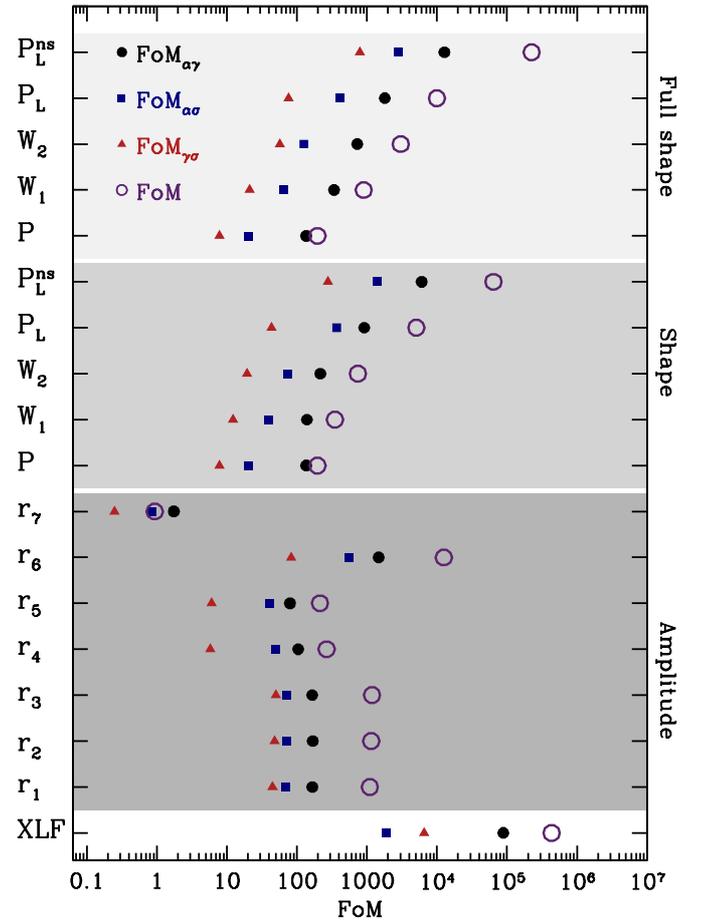}
\caption{Figures-of-merit obtained from the $1\sigma$ error ellipses of each pair of parameters of the scaling relation, taken from the different observables $\mu_{j}$ defined in Eq.~(\ref{muq}). By FoM we denote the figure-of-merit computed as the volume of the 3-dimensional $68\%$ error ellipsoid. To facilitate the understanding of this figure, the dark gray, gray and light gray areas denote observables accounting for the amplitude, the shape and the full shape of power spectra respectively. Results are shown in real space. Noticeably, the luminosity power spectrum $P_{L}$ provides more information regarding the scaling relation than the luminosity-weighted spectra, and display a total FoM that is $\sim 10^{2}$ times that of the unweighted power spectrum. We show the estimates derived from the luminosity power spectrum $P^{\rm ns}_{L}(k)$, in order to emphasize how high FoM can be obtained when no shot-noise correction is spuriously implemented in the measurements.}\label{rex}
\end{figure}

Figure~\ref{red22} shows the error ellipses from the analysis of the full shape of the power spectra $\hat{P}(k)$, $\hat{\maw}_{1,2}(k)$, $\hat{P}_{L}(k)$ and $\hat{P}^{\rm ns}_{L}(k)$, both in real and in redshift space, with Fourier modes in the range $0.02<k/(h\,{\rm Mpc}^{-1})<0.2$. The left hand panel in that figure corresponds to the results obtained after marginalizing the covariance matrix of each observable with respect to an overall amplitude, while the right hand panel shows results obtained after using the information encoded in the full shape. This figure confirms the intuitive prediction that more information (smaller error ellipses) is gained when the full shape is used in the analysis. 
We complement the information shown in Fig.~\ref{red22} with the information shown in Fig.~\ref{kmax} and~\ref{rex}. From these we conclude the following.
\begin{itemize}

\item The marked power spectra $\hat{\maw}_{1,2}(k)$ generate higher FoM compared to $P(k)$, due mainly to their sensitivity to the slope of the scaling relation $\gamma$. These three spectra are almost equally sensitive to the intrinsic scatter $\sigma$ and the amplitude $\alpha$.
\item A gain in information content on the scaling relation is obtained from the luminosity power spectrum, with FoM increased approximately by an order of magnitude over what is obtained with $\maw_{2}$, or, $\sim 40$ times below what characterizes the information encoded in the XLF. 

\item Figure~\ref{kmax} shows the behavior of the FoM related to the full shape of the observables $\hat{P}(k)$, $\hat{\maw}_{1,2}(k)$, and $\hat{P}_{L}(k)$, as a function of the maximum wavenumber used in the analysis, $k_{\rm max}$. We see that the ratio of the FoM contained in the luminosity power spectrum $P_{L}(k)$ to that of the unweighted power spectrum $P(k)$ ranges from a factor $10^{2}$ with $k_{\rm max}\sim 0.07 h{\rm Mpc}^{-1}$ to $\sim 20$ with $k_{\rm max}
\sim 0.2 h{\rm Mpc}^{-1}$. The trend of the different FoM with the maximum wavenumber is as expected; i.e, the higher the number of Fourier modes, the higher the content of information in the Fisher matrix.

\item To emphasize the relevance of the shot-noise correction, we repeated our analysis using the shot-noise uncorrected estimates of the different power spectra. As a result, the FoM obtained are higher than those presented in Fig.~\ref{rex}, as has been explicitly portrayed for the case of the shot-noise uncorrected luminosity power spectrum (denoted by $P^{\rm ns}_{L}(k)$). The ratios between the FoM obtained from the uncorrected estimates to those derived from the corrected ones (using their full shape as probes) are $\sim 3$, $21$, and $23$ for $P,\maw_{2}$, and $P_{L}$ respectively, illustrating  how sensitive the clustering signal of galaxy clusters to the shot-noise correction is. Furthermore, the total FoM associated to $P^{\rm ns}_{L}(k)$ is a factor $\sim 2.5 $ smaller than the corresponding value obtained from the XLF, which is noticeable if we compare it to the factors $\sim 15$ and $\sim  2\times 10^{3}$ derived from $\maw_{2}(k)$ and $P(k)$ respectively \footnote{The Fisher matrix of $P_{L}$ \emph{is not} the sum of the Fisher matrix of the weighted spectra.}. The shot-noise correction (either Poisson-like or scale-dependent) is always demanded, if not in the measurement, then in the model adopted for interpreting the clustering signal. Not doing that can lead to biased interpretations of the observed power spectra, which degenerate into spurious tight constraints on the parameters of interest.
\end{itemize}

In summary, Fig.~\ref{rex} attempts to answer the question posed by the title of this paper. The improvement in the FoM obtained from the luminosity power spectrum with respect to the standard probes (i.e., unweighted clustering analysis) represents the main result of our analysis. 

\section{Discussion and conclusions}\label{sec:conc}
In this paper we aimed to explore the capability of the so-called marked statistics of galaxy clusters to retrieve information on the links between cluster masses and cluster X-ray luminosities. To this end, we assigned X-ray luminosities to dark matter halos in a suite of $N$-body simulations and measured luminosity-weighted power spectra $\hat{\maw}_{1,2}(k)$ and luminosity power spectrum $\hat{P}_{L}(k)$, as defined in Sect.~\ref{sec:meas}. The luminosity power spectrum has been defined such that it directly measures the clustering of X-ray luminosity. The definition of the luminosity power spectrum can be extrapolated to other probes of two-point statistics, such as the angular power spectrum. This can be especially suitable in cases where no redshift-information is available and clusters are identified by a redshift independent intrinsic property, such as the SZ signal \citep[e.g.,][]{2013arXiv1303.5089P}.

We extracted the information related to the scaling relation by dissecting the clustering signal in three different parts: first, we isolated the large-scale information on the power spectra measured in different luminosity bins, and measured estimates of luminosity bias. Second, we explored the information encoded in the shape of the clustering signal, and third, we considered their full shape (i.e., shape and amplitude). This yields a total of twelve observables. By carrying out a Fisher matrix analysis, we quantified the amount of information regarding the cluster scaling relation from each of these observables through the definition of a FoM. In terms of the \emph{amplitude}, we showed that the information encoded in the estimates of cluster luminosity bias measured from the unweighted power spectrum $P(k)$ is close to what is obtained from the bias of the weighted spectra $\maw(k)$, so that no sensitive gain of information is achieved. However, a relevant increase in the information content is obtained when estimates of relative bias are extracted from the luminosity power spectrum, reaching FoM that are only one order of magnitude below the values derived from the XLF analysis. Similarly, when exploring the information content within the full shape of the power spectra, we found that the information associated to $P_{L}(k)$ overcomes that of the spectra $\maw_{1,2}(k)$ and $P(k)$: the resulting FoM of the $P_{L}$ is a factor of $\sim 100$ ($\sim 10$) with respect to that of the unweighted (weighted) power spectrum. \emph{The gain in information obtained from the estimates of the luminosity power spectrum with respect to the unweighted analysis} represents the main result of this paper. 

The results presented show, though, that the XLF still encodes more astrophysical information than our best clustering probe, the luminosity power spectrum. This implies that the contribution of the latter in the constraints of astrophysical parameters within a joint XLF-clustering analysis could be subdominant, at least in the case where the cosmological sector is fixed. In a more realistic scenario, though, \citet{pillepich} show that the combination of cluster number counts and the standard (i.e, unweighted) angular cluster power spectrum for the \emph{eROSITA} experiment can set constraints that are tighter than those derived from the number counts alone. Nevertheless, these authors emphasize that this trend can be reversed after variations in the selection function of the sample in their study. A direct comparison between our approach and the \citet{pillepich} analysis is not applicable, since, as we already pointed out, we are not interested in assessing the ability of a certain model to explain a given experiment, and also because we have kept the cosmological parameters fixed, thereby ignoring all possible correlations between the cosmological and the astrophysical sector. In view of our results, it can be expected that the joint analysis of number counts and some \emph{X-property} angular power spectrum can contribute to improving the FoM in the work of \citet{pillepich}. This is currently under study.

An obvious question arises, as to the feasibility of the clustering signal to set precise constraints on cluster scaling relations in future galaxy-cluster surveys. From the observational point of view, the volume of the $N$-body simulations we used to create cluster catalogs mimics that of a full-sky complete sample to a maximum redshift of $z\sim 0.3$. The resulting mean number density resembles what is expected from the \emph{eROSITA} experiment \citep{pillepich}. Therefore the precision to which \emph{only} the scaling relation could be constrained from these forthcoming samples \--when redshifts are available and using $P_{L}(k)$ \--- can be of the same order as that obtained from our analysis. 
On the other hand, the volumes that will be probed by the forthcoming surveys will generate statistical errors on the estimates of power spectra that might be compatible to the systematic errors present when modeling clustering. This situation is especially noticeable on small scales \--- probed by a large number of modes\--- where the clustering signal is dominated by the nonlinear evolution of the matter density field, the scale-dependent halo-mass bias, the halo exclusion, and ultimately baryonic effects. 

In Fig.~\ref{rex} we displayed our results such that relevant FoM are obtained, especially for the luminosity power spectrum, when we include Fourier modes up to a scale of $k\sim 0.2\,h\,{\rm Mpc}^{-1}$ in our analysis. From a theoretical perspective, extending the analysis to such values of wavenumbers might be a point of concern. On these scales, highly complex processes (e.g., nonlinear evolution of the matter density field, scale-dependent halo biasing and halo exclusion effect) become non-negligible. In the past two decades, major progress in the understanding of some of these processes has been accomplished. In that regard, there are several attempts to describe the nonlinear evolution of dark matter, either based on numerical fits to $N$-body simulations \cite[e.g.,][]{1996MNRAS.280L..19P,2003MNRAS.341.1311S} or on theoretical predictions \cite[e.g.,][]{2002PhR...367....1B,2006PhRvD..73f3519C,2007PhRvD..75d3514M,2011PhRvD..83h3518M}. The link between dark matter and halos \---the halo-mass bias\--- has also been an active area of research, regarding its nature and fitting formulae for its implementation \citep[see, e.g.,][for a recent review of this subject]{2013P}. Even if these theoretical predictions accurately describe the results from $N$-body simulations, baryonic effects within galaxy clusters (e.g., radiative cooling, star formation, feedback mechanism due to AGN and supernova) can severely spoil the expectations of modeling the observed clustering signal to the accuracy achieved by state-of-the art cosmological simulations of dark matter. Indeed, such effects have been shown to substantially modify the abundance and clustering of clusters with respect to pure dark matter-based predictions \citep[e.g.,][]{stanek_bar,rudd,daalen,cui,2013JCAP...04..022B}. In view of the plethora of models accounting for the baryonic effects that shape the intracluster medium and the small scales (astrophysical, instead of cosmological) where these effects take place, precise modeling of the abundance and clustering of galaxy clusters based on $N$-body simulations is still far from being achieved.
Even though it is beyond the scope of this paper to discuss the impact of the systematic effects present on the modeling of the quantities mentioned above, we recognize their relevance in extracting accurate information regarding the parameters describing the observed clustering pattern of galaxy clusters (especially using the luminosity power spectrum) from forthcoming galaxy and galaxy cluster surveys such as \emph{eROSITA}, DES \citep{abbott}, or Euclid \citep{laureijs}. We notice that the only model that we have explicitly implemented in our analysis makes the assumption that the halo distribution behaves like a Poisson point process, which is a debatable hypothesis \citep[see, e.g.,][]{2002MNRAS.333..730C,2007PhRvD..75f3512S}. With this assumption, the shot-noise correction of the halo power spectrum is represented by subtracting a white noise from the raw power spectrum. However, even simple toy models assuming spherical halos reveal a scale-dependent shot noise. We discuss this briefly in Appendix~\ref{exclusion}.

The weighted two-point statistics proves itself to be an interesting tool for characterizing the explicit dependence of the clustering of galaxy clusters on intrinsic properties. The implementation of these statistics, either in Fourier space (as we have shown in this paper) or in configuration space \citep[see, e.g.,][]{2005MNRAS.364..796S}, can contribute setting tight constraints on the parameters that characterize the physics of the intra-cluster medium, interestingly intertwining cosmology and astrophysics on the galaxy cluster-scale.

\paragraph*{Acknowledgments}
{\small I am grateful to Ra\'ul Angulo for making the L-BASICC II simulations available. I thank Cristiano Porciani and Nina Roth for useful discussions and suggestions, as well as the anonymous referee for comments that helped improve the presentation and content of the manuscript. I acknowledge support through the SFB-Transregio 33 ``The Dark Universe'' by the Deutsche Forschungsgemeinschaft (DFG).}

\bibliographystyle{aa}        
\bibliography{refs}

\begin{appendix}
\section{The cluster-mass bias}\label{apl}
The ratios $r_{\nu}(L)$ defined in Sect.~\ref{sec:amp} can be predicted under the assumption that, on large scales, dark matter halos of a given mass $M$ are biased tracers of the underlying dark matter distribution. In the so-called local bias model, to zeroth order in perturbation theory \citep[e.g.,][]{2002PhR...372....1C}, this is expressed as $P(k;M)=b^{2}(M)P_{\rm mat}(k)$, where $P_{\rm mat}(k)$ denotes the power spectrum of the dark matter and $b(M)$ is a scale-independent halo-mass bias. Given a halo mass function $n(M)\dd M$ (i.e., the number of dark matter halos with masses between $M$ and $M+\dd M$ per unit comoving volume) and a selection function $\hat{\phi}(M,L)$ (i.e., the probability of having a cluster with mass $M$ given some selection criteria), the real-space marked power spectra can be written as $\maw_{\nu}(k;L)=(\tilde{b}_{w}(L)/\tilde{b}(L))^{\nu}P(k;L)$ \citep[e.g.,][]{2002PhR...372....1C,2005MNRAS.364..796S}, where 
\be\label{bi2}
\tilde{b}_{w}(L)\equiv \frac{\int_{0}^{\infty} n(M)\hat{\phi}(M,L)b(M)\langle L |M \rangle \dd M}{\int _{0}^{\infty} n(M)\hat{\phi}(M,L)\langle L |M \rangle \dd M},
\ee
and
\be\label{bb}
\tilde{b}(L) \equiv\frac{\int_{0}^{\infty} n(M)\hat{\phi}(M,L)b(M)\dd M}{\int_{0}^{\infty} n(M)\hat{\phi}(M,L)\dd M},
\ee
is the effective cluster-matter bias. For a given bin of X-ray luminosity, the selection function $\phi(M,L)$ can be expressed as the average of the scaling relation $\mapp(L|M)$ given by Eq.~(\ref{sr}) in that luminosity bin. The ratios $r_{1,2,3}(L)$ in real space can therefore be interpreted as estimates of $b(L)/b(L_{\rm ref})$, $b_{w}(L)b(L)/(b(L_{\rm ref})b_{w}(L_{\rm ref}))$, and $b_{w}(L)/b_{w}(L_{\rm ref})$ respectively. Also, the information contained in the ratio $r_{4}(L)$ is the same as contained in the ratio $r_{5} (L)$, since these are estimates of the ratio $(b_{w}(L)/b(L))^{1/2}$ and $b_{w}(L)/b(L)$ respectively. Accordingly, the luminosity power spectrum can be written as 
\be\label{plt}
P_{L}(k;L)=\lp \frac{\tilde{b}_{w}(L)}{\tilde{b}(L)}-1\rp^{2} P(k;L),
\ee
from which the predictions for the ratios $r_{6,7}(L)$ can be readily obtained. 
According to  Eq.~(\ref{sr}), the moments of the luminosity $L$ are linked to the scaling relation via $\langle L^{n}|M\rangle={\rm exp}\lp n\langle \ell|M\rangle+\frac{n^{2}}{2}\sigma^{2}\rp$. Therefore, the bias $\tilde{b}_{w}$ is directly sensitive to the mean of the scaling relation, yet indirectly (only through $\hat{\phi}(M,L)$) to the intrinsic scatter.

\begin{figure}
\center
\includegraphics[width=8cm, angle=0]{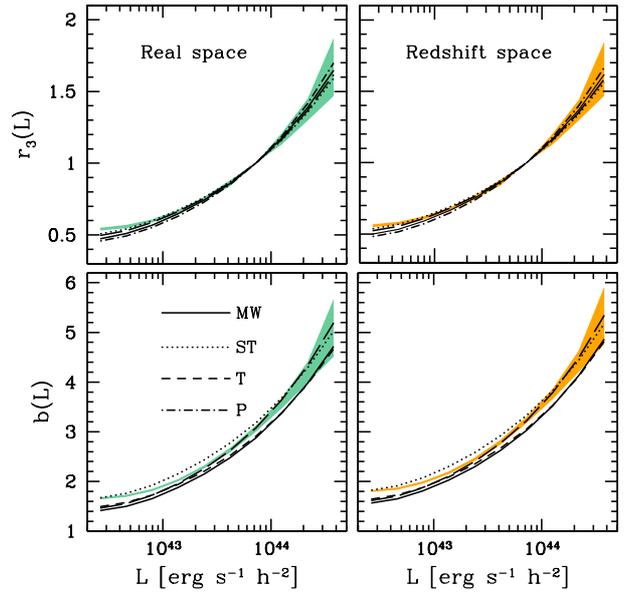}
\caption{The ratio $r_{3}(L)$ (top panels) defined in Eq.~(\ref{r1}), and the luminosity bias $b(L)$ (bottom panels) obtained in the range $0.02\leq k/(h\,{\rm Mpc}^{-1})\leq 0.08$. Results are shown in real (right panel) and redshift (left panel) space. For readability, we only show the standard deviation obtained from the $N$-body simulations with shaded regions. The lines represent the predictions presented in Sect.~\ref{sec:amp} using expressions for the halo-mass bias as reported by different authors: \cite[][MW]{1996MNRAS.282..347M}, \cite[][ST]{1999MNRAS.308..119S}, \cite[][T]{2010ApJ...724..878T}, and \cite[][P]{pillepich_nm}.}
\label{rnu}
\end{figure}

The redshift-space estimates of the ratios $r_{\nu}(L)$ can be obtained in a similar way. Under the plane-parallel approximation, the large-scale signal of the redshift-space cluster power spectrum $P^{s}(k;L)$ can be described by the so-called Kaiser effect \citep[e.g.,][]{1987MNRAS.227....1K,1998ASSL..231..185H} $P^{s}(k;L)=\lp 1+2\beta/3+\beta^{2}/5\rp P(k;L)$, where $\beta\equiv f/\tilde{b}(L)$ and $f\equiv {\rm d}\ln D(a)/{\rm d}\ln a$ is the growth index ($D(a)$ represents the growth factor) \citep[e.g.,][]{1980lssu.book.....P}. Given the cosmology and redshift output of the L-BASICC simulations that we used, $f=0.44$. The Kaiser effect can be generalized to the marked power spectra \citep[e.g.,][]{2006MNRAS.369...68S} as
\be\label{w1s}
\maw^{s}_{1}(k;L)=\frac{\tilde{b}_{w}(L)}{\tilde{b}(L)}\lp1+\frac{1}{3}\lp\beta+\beta_{w}\rp+\frac{1}{5}\beta\beta_{w}\rp P(k;L),
\ee
and
\be\label{w2s}
\maw^{s}_{2}(k;L)=\frac{\tilde{b}^{2}_{w}(L)}{\tilde{b}^{2}(L)}\lp1+\frac{2}{3}\beta_{w}+\frac{1}{5}\beta_{w}^{2}\rp P(k;L),
\ee
where $\beta_{w}\equiv f/\tilde{b}_{w}(L)$. We have checked whether these expressions describe the ratios $r_{\nu}(L)$. To this end, the halo abundance $n(M)$ is taken to be described by the fitting formulae of \cite{jenkins}, which is suitable for simulations such as the L-BASICC. A crucial step is to choose the halo-mass bias $b(M)$. In Fig.~\ref{rnu} we show predictions for the ratio $r_{3}(L)$, obtained using some examples of prescriptions for this quantity: \cite{1996MNRAS.282..347M}, \cite{1999MNRAS.308..119S}, \cite{2010ApJ...724..878T}. and \cite{pillepich_nm}. To witness the performance of these prescriptions, the bottom panels in Fig.~\ref{rnu} show the luminosity bias obtained similar to what is described by Eq.~(\ref{r1}), using the estimates of the matter power spectrum of the L-BASICC II simulation. This figure shows that $i)$ as established by a number of studies, a scale-independent halo-mass bias is a fair modeling of the cluster power spectrum on large scales. $ii)$ The Kaiser effect is a good description of the redshift-space power spectra, at least within the range of masses and scales probed by our analysis \citep[see, e.g.,][for a broader discussion on this subject]{2012MNRAS.427.2420B} $iii)$ When assessing the ability of a model to retrieve either cosmological or astrophysical information from one or two-point statistics, extreme caution is required in view of the discrepancies observed between models and the simulations. In particular, more accurate models of halo-mass function and halo-mass bias are demanded, given the small statistical errors expected from forthcoming surveys \citep[see, e.g.,][]{2010PhRvD..81h3509C, 2010ApJ...713..856W, 2012arXiv1211.6434S} with volumes comparable to that of the L-BASICC simulation. The differences between the different fitting formulae presented in  Fig.~\ref{rnu} can be caused by several effects, namely, the difference cosmological models and/or parameters used in the $N$-body simulations used to fit each of them, the characteristics of the halo-identification algorithm (which introduces systematic effects both in the mass function and the halo-mass bias), and the way the biases are measured (i.e, either from the correlation function, the power spectrum, or by means of count-in cells experiments). It is beyond the scope of this work to analyze these differences in detail; however, we note that different fitting formulae can provide a good description of the measured luminosity bias at different ranges of X-ray luminosity. In particular, the results from \cite{2010ApJ...724..878T} generate a fair description for luminosities above $\sim 3\times 10^{43}h^{-2}{\rm erg}s^{-1}$. Lower luminosities correspond to halos defined by a relatively low number of dark matter particles, where resolution effects can be relevant. Finally, $iv)$ the discrepancies between the different prescription of halo-mass bias are slightly diminished when we work with estimates of relative biases instead of absolute biases.

\section{Halo exclusion}\label{exclusion}

\begin{figure}
\center
\includegraphics[width=9cm, angle=0]{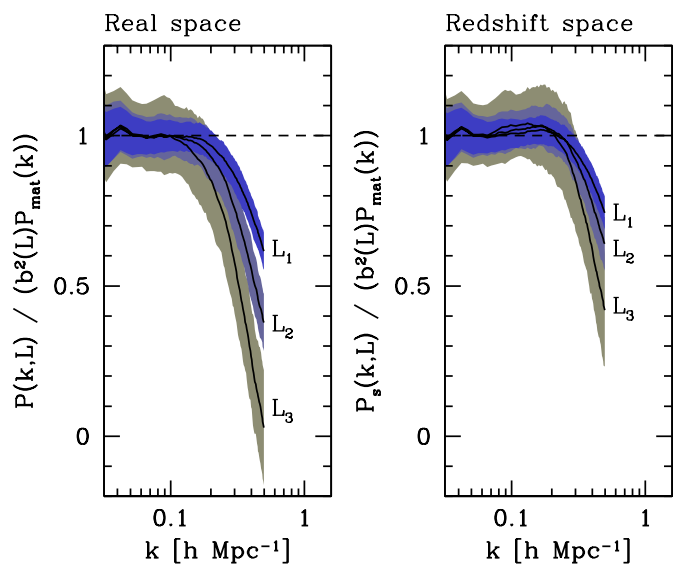}
\caption{Halo exclusion: the plot shows the ratio between the measured cluster power spectrum described in \S~\ref{sec:meas} and the expected linear cluster power spectrum $b^2(L)P_{\rm mat}(k)$, for clusters in four different bins of X-ray luminosity characterized by a luminosity $L_{i}$, with $L_{3}>L_{2}>L_{1}$. The luminosity bias $b^{2}(L)$ is that measured from the simulations. The shaded regions and the solid line represent the standard deviation and the mean, respectively, obtained from the $N$-body simulations.}
\label{ex}
\end{figure}
Dark matter halos are not point-like objects. As a consequence, the idea that their spatial distribution can be described as a Poisson-point process drawn from a realization of an underlying continuous field with a positive correlation function \citep{1980lssu.book.....P} is, strictly speaking, not a realistic assumption. So far, a statistical description of halo distribution that takes their finite size into account is not fully accomplished. Instead, simple geometrical approaches have been developed to empirically model the exclusion effect within the context of the two-point correlation function. Here we briefly illustrate how this model works. Assuming that we can assign a radius to each spherical halo, the correlation function can be written as \citep[e.g.,][]{2002ApJ...565...24P} 
\be\label{ex-geo}
\xi_{\rm h}(r;M,M')=
\begin{cases}
  b(M)b(M')B^{2}(r)\xi_{\rm mat}(r) & r >  \bar{R}(M,M'),\\ 
 -1 & r\leq  \bar{R}(M,M'),
\end{cases}
\ee
where $\bar{R}(M,M')\equiv \langle R|M\rangle+\langle R|M'\rangle$, with $\langle R|M\rangle$ denoting the expected radius of a cluster with mass $M$. In this expression, $b(M)$ denotes the dark-matter halo scale-independent bias, $B(r)$ denotes a possible scale-dependency in the halo-matter bias \cite[e.g.,][]{2005ApJ...631...41T}, while $\xi_{\rm mat}(r)$ is the full nonlinear matter correlation function. For halos with masses within an infinitesimally narrow range, this expression predicts a sharp transition towards $\xi_{\rm h}(r)=-1$ at a scale equal to twice the radius of the halo. This transition becomes smoother when halos with different masses (and thus sizes) are included. In Fourier space, the exclusion is translated to a lack of power on small scales, which goes counter to the effect of the nonlinear clustering. Since the exclusion effect is more evident when more massive halos are considered, such lack of power displays a clear trend with the characteristic mass (or X-ray luminosity) of the sample. The halo power spectrum from Eq.~(\ref{ex-geo}) can be separated into three components:
\be
P_{\rm h}(k;M,M')=P^{0}_{\rm h}(k;M,M')-\frac{4\pi \bar{R}^{2}}{k}j_{1}(k\bar{R})+P_{1}(k,M,M').
\ee
The first term $P^{0}_{\rm h}(k;M,M')=b(M)b(M')P_{\rm mat}(k)$ is the halo power spectrum with a scale-independent bias. The second term is the Fourier transform of the $-1$ in Eq.~(\ref{ex-geo}), where $j_{1}(x)$ denotes the spherical Bessel function of first order. The third contribution can be written as
\be\label{a1}
P_{1}(k;M,M')=\frac{2}{\pi}\int_{0}^{\infty} \tilde{k}^{2}P^{0}_{\rm h}(\tilde{k};M,M')G(k,\tilde{k},\bar{R})\,\dd \tilde{k},
\ee
where the kernel $G(k,\tilde{k})$ is given by $G(k,\tilde{k})=G_{1}(k,\tilde{k},\bar{R})+G_{2}(k,\tilde{k},\bar{R})$, with
\be
G_{1}(k,\tilde{k},\bar{R})= \int_{\tilde{R}}^{\infty}\,r^{2}j_{0}(kr)j_{0}(\tilde{k}r)(B^{2}(r)-1)\,\dd r,
\ee
and
\be
G_{2}(k,\tilde{k},\bar{R})=\frac{k\sin(\tilde{k}\bar{R})\cos(k\bar{R})-\tilde{k}\sin(k\bar{R})\cos(\tilde{k}\bar{R})}{\tilde{k}(\tilde{k}^{2}-k^{2})}.
\ee
In the limit $\bar{R}\to 0$ (i.e., no halo exclusion), both the second term on the right hand side of Eq.~\ref{a1} and the kernel $G_{2}$ go to zero. In that case a nonlinear contribution to the cluster power spectrum relies on the behavior of the scale-dependent halo-mass bias $B(r)$. In the limit of an homogeneous distribution, we end up with a power spectrum of the form $-4\pi \bar{R}^{2}j_{1}(k\bar{R})/k$. Thus, to obtain an unbiased estimation of the cluster power spectrum (of spherically symmetric nonoverlapping clusters), this last term would need to be subtracted from the raw estimates (as in Eq.~\ref{pp}), together with the white shot noise $1/\bar{n}$. The combination of these two effects can be regarded as scale-dependent shot-noise.

Figure~\ref{ex} depicts the halo exclusion effect as measured from the ensemble of halos presented in \S~\ref{sec:sim}. As pointed out above, the strength of the effect scales with the luminosities (or masses) of the objects considered in the analysis. Therefore, this signal can be used to retrieve information on the underlying scaling relation (e.g., mass-X-ray luminosity when analyzing cluster samples or the mass-number of hosted galaxies when analyzing a galaxy redshift survey)  \citep[e.g.,][]{2002ApJ...565...24P}.  Finally, the exclusion effect is attenuated when observed in redshift space, simply because pairs of halos are observed to be closer due to their peculiar velocities. For instance, exploring the power spectrum obtained from the full luminosity sample shows that in real space exclusion sets in at $k\sim 0.2 \,h\,{\rm Mpc}^{-1}$, while this value shifts to $\sim 0.3\,h\,{\rm Mpc}^{-1}$ in redshift space.

\end{appendix}

\end{document}